\documentstyle[12pt]{article}
\begin{document}

\newcommand{\beq}{\begin{equation}}
\newcommand{\eeq}{\end{equation}}
\newcommand{\beqn}{\begin{eqnarray}}
\newcommand{\eeqn}{\end{eqnarray}}
\newcommand{\beqns}{\begin{eqnarray*}}
\newcommand{\eeqns}{\end{eqnarray*}}
\newcommand{\nn}{\nonumber}

\def\Ac{{\cal A}}
\def\Bc{{\cal B}}
\def\O{{\cal O}}
\def\F{{\cal F}}
\def\Pc{{\cal P}}

\def\qb{\bar{q}}
\def\Qb{\bar{Q}}

\def\to{\rightarrow}
\newcommand{\lra}{\leftrightarrow}
\newcommand{\la}{\langle}
\newcommand{\ra}{\rangle}
\def\A#1#2{\la#1#2\ra}
\def\B#1#2{[#1#2]}
\def\s#1#2{s_{#1#2}}
\def\h#1#2#3#4{\la#1#2#3#4\ra}
\def\P#1#2{{\cal P}_{#1#2}}
\def\L#1#2{\left\{#2\right\}_{#1}}
\newcommand{\Lmin}{{\rm L}}
\def\Li{{\rm Li_2}}
\def\ms{$\overline{{\rm MS}}$}

\newcommand{\y}{\gamma}
\newcommand{\yf}{\gamma_5}
\newcommand{\yh}{\hat{\gamma}}
\newcommand{\yt}{\tilde{\gamma}}
\newcommand{\yb}{\bar{\gamma}}
\newcommand{\g}{g}
\newcommand{\gh}{\hat{g}}
\newcommand{\gt}{\tilde{g}}

\newcommand{\al}{\alpha}
\newcommand{\be}{\beta}
\newcommand{\del}{\delta}
\newcommand{\eps}{\epsilon}
\newcommand{\ve}{\varepsilon}

\newcommand{\kt}{\tilde{k}}
\newcommand{\Pp}{{\cal P}_+}
\newcommand{\Pm}{{\cal P}_-}
\newcommand{\Ppm}{{\cal P}_\pm}
\newcommand{\Pmp}{{\cal P}_\mp}
\def\slash#1{/ \! \! \!#1}

\newcommand{\Nf}{N_F}
\newcommand{\Nc}{N_c}
\newcommand{\cg}{c_\Gamma}
\newcommand{\eL}{[e]}
\newcommand{\eLL}{(e^2)_{\rm Loop}}
\newcommand{\tr}{\rm tr}
\newcommand{\inteq}{{ {\int} \! \! \! \! \! =} \;}
\newcommand{\intsim}{{ {\int} \! \! \! \! \! \sim} \;}

\begin{titlepage}
\begin{flushright}
ETH-TH/95-19\\
June 1995 \\
\end{flushright}
\vskip .5in
\begin{center}
{\Large\bf One-loop corrections to five-parton amplitudes with external
  photons  }
\footnote{Work supported in part by the Schweizerischer
Nationalfonds}\\
\vskip 1cm
{\large Adrian Signer} \\
\vskip 0.2cm
Theoretical Physics, ETH, \\
Z\"urich, Switzerland  \\
\vskip 1cm
\end{center}

\begin{abstract}
\noindent
Recently the calculations of all five-parton one-loop QCD amplitudes have
been completed. In this letter we describe how to get the corresponding
amplitudes with one gluon replaced by a photon and we give the explicit
results for the process $0 \to 2q 2Q 1\y$.
\end{abstract}

\end{titlepage}

\setcounter{footnote}{0}

\section{Introduction}

The study of photoproduction of 3+1 jets at HERA,
as well as the study of direct photon
production in connection with two jets at hadron colliders, requires the
knowledge of five-parton amplitudes with external photons. Since by now all
one-loop five-parton QCD amplitudes are known \cite{BDK5g,KST4q1g,BDK2q3g},
it is natural to exploit these results in order to get the desired
amplitudes.

For amplitudes with no external quarks the photon can couple  through a
quark loop only. As a result, the corresponding tree-level amplitudes
vanish. We therefore concentrate on the process with external quarks.

After the setup of the notation we describe how to get the amplitudes for
$0 \to 2q 2g 1\y$ and $0 \to 2q 1g 2\y$
out of $0 \to 2q 3g$ and we give the complete results for
$0 \to 2q 2Q 1\y$. Of course, it is also possible to
get the amplitudes for $\ 0 \to 2q 3 \y $ and the pure QED
processes out of the corresponding QCD amplitudes \cite{BDK2q3g,thesis},
but these processes are phenomenological less important.
The construction of  photonic one-loop amplitudes for one external quark
pair has also been discussed in  \cite{BDK2q3g}.

The quarks (antiquarks) are
denoted by $q$ and $Q$ $\; (\qb \ \rm{and} \ \Qb)$, the photon by $\y$
and the gluons are labeled  from 1 to $n$.
These labels may be used for the color
index or the momentum of the corresponding particle. The dependence on the
helicities  $h_i$ of the particles  is often suppressed

Throughout we use the normalization $Tr(T^a T^b) = \del^{a b}$, where $T^a$
are the $SU(3)$ generators in the fundamental
representation. Furthermore, we use the following compact notation for
the color part:
\beqn
(i\ldots j)_{q \qb} &\equiv&  (T^{g_i} \ldots T^{g_j})_{q \qb}
\label{colnot1} \\
(i\ldots j) &\equiv& \mbox{Tr} (T^{g_i} \ldots T^{g_j}) \label{colnot2}
\eeqn
After replacing the structure constants $f^{abc}$ with $-i/\sqrt{2}
([a,b] c)$ and using the Fierz identity, the color part of every Feynman
diagram can be written as a linear combination of terms of the form given
in eqs. (\ref{colnot1}) and (\ref{colnot2}). The coefficient of an amplitude
proportional to such a color factor is called a subamplitude.

For the presentation of the (unrenormalized) amplitudes we follow closely
the conventions in ref. \cite{KST4q1g}.  The amplitudes will be expressed
as functions of dot products, spinor inner products and a set of auxiliary
functions $ ( \L x {^{ij}_{kl}}, \ \h i j k l, \ \P i j \; {\rm and} \;
\F(i,j,k) ) $. All these functions as well as the ubiquitous prefactor
$c_\Gamma$ are defined in ref. \cite{KST4q1g}.

\section{One external quark pair}

Consider the  process $0\to 2q 3g$ where one gluon, say $g_3$, is replaced
by a photon. The only
differences of  the two processes are the coupling constant and the
color part. First of all we have to replace $g \to e_q$, where $e_q$ is the
electric charge of the quark to which the photon couples,
and secondly we have to make the transition $T^{g_3} \to \sqrt{2} $.
\footnote{The factor $\sqrt{2}$ appears because of the chosen normalization
  of the $T^i$}

Although both steps seem to be rather trivial, they may cause some
complications. The former, because the photon may couple to quarks with
different electric charge and the latter, because we have to calculate
additional color subamplitudes which belong to color structures
involving a
trace ${\rm Tr}(T^{g_3})$. Contrary to the gluonic process this trace is not
vanishing any longer after the transition $T^{g_3} \to \sqrt{2} $.

As long as we have only one external quark and the corresponding antiquark,
the difficulty concerning the
replacement $g\to e_q$ can be handled as follows: First of all, for
diagrams without a fermion loop there is no problem, we always have
$g \to e_q$, where $e_q$ is the electric charge of the
external quark.
For diagrams with a fermion loop we have to separate the
diagrams where the  photon couples to an internal fermion
loop (called loop-coupled diagrams)
from those where the photon couples to the external quark (externally-coupled
diagrams).
In the former case we have to replace $g \to \sum e_Q \equiv
[e]$ where the
sum extends over all fermions $Q$ in the loop whereas in
the latter case we have the
usual replacement $g \to e_q$.

Since the  subamplitudes $0\to 2q3g$ are needed for the
presentation of the results we review
the color structure of the helicity amplitudes of this process.
At tree level it is given by
\beq
\Ac^{(0)}(q,1,2,3,\qb) = g^3 \sum_{P(123)} (123)_{q\qb}
 \ b_{123}^{(0)} (q,1,2,3,\qb)
\eeq
where $P(1\ldots n)$ denotes all permutations of the elements $ 1 \ldots n$.
As mentioned above, we suppress the helicity dependence of the amplitudes in
the notation.
At one loop the decomposition reads:
\begin{eqnarray}
\label{oneloopB}
&& \Ac^{(1)}(q,1,2,3,\qb) =     g^3\left({g\over 4\pi}\right)^2 \Bigg[
\sum_{P(1,2,3)}  (1 2 3)_{q\qb} \ b^{(1)}_{123} (q,1,2,3,\qb) \\
&& \qquad \qquad + \ \sum_{i=1,2,3} \left(
(i)_{q\qb} (kl) \ b^{(1)}_i (q,1,2,3,\qb)
+ (i) \ (kl)_{q \qb} \ b^{(1)}_{i;kl} (q,1,2,3,\qb) \right) \nn \\
&&  \qquad \qquad + \ (q\qb) \ (123) \ b^{(1)}_q (q,1,2,3,\qb)
+ \ (q\qb) \ (321) \ b^{(1)}_{\qb} (q,1,2,3,\qb)
\Bigg] \nn
\end{eqnarray}
In this equation $k$ and $l$ are the two indices from the set
$\{1,2,3\}\setminus i$ for a given $i\in \{1,2,3\}$.
The subamplitudes $ b^{(1)}_{i;kl}$ are not needed for $0 \to 2q3g$ since
the corresponding color part vanishes. However, as we shall see, these
subamplitudes have to be known for the construction of the photonic
processes. All subamplitudes can either be extracted from
ref. \cite{BDK2q3g} or they are explicitly given in ref. \cite{thesis}.

If we replace $g_3 \to \y$, the color decomposition at tree level is
\beq
\Ac^{(0)}(q,1,2,\y,\qb) = g^2  \sqrt{2} e_q \sum_{P(1,2)} (1 2)_{q \qb} \
 b_{12}^{1\y,(0)}(q,1,2,\y,\qb)
\eeq
while at one-loop we have
\beqn
\lefteqn{\Ac^{(1)}(q,1,2,\y,\qb) = } \\ \nn
&&g^2  \sqrt{2} e_q \left({g\over 4\pi}\right)^2 \left[
 \sum_{P(1,2)} (12)_{q \qb} \ b_{12}^{1\y,(1)} (q,1,2,\y,\qb) +
 (q \qb) (12) \ b_{\tr}^{1\y,(1)} (q,1,2,\y,\qb) \right] \label{coldec2q2g1y}
\eeqn
Note that according to the discussion above the subamplitudes may contain
terms proportional to $[e]/e_q $.

The tree-level results can be obtained immediately with
\beq
b_{12}^{1 \y,(0)}(q,1,2,\y,\qb)  =  b_{123}^{(0)}(q,1,2,\y,\qb) +
b_{132}^{(0)}(q,1,2,\y,\qb) + b_{312}^{(0)}(q,1,2,\y,\qb)
\eeq

For the subamplitude $b_{\tr}^{1\y,(1)}$ the
externally-coupled diagrams
 cancel each other, therefore, we have no
contributions proportional to $e_q \Nf$. This fact enables us to get the
subamplitudes $b_{\tr}^{1\y,(1)}$ out of those of the process $0 \to
2q3g$. To do so we merely have to list  all color factors of
$0 \to 2q3g$ which, after the transition $g_3 \to \y$,
yield the color part $(12)$. These are $(123), (321)$ and $(3)_{\qb q}(12)$.
Adding up the corresponding subamplitudes yields the desired result
\beqn
\lefteqn{b_{\tr}^{1\y,(1)}(q,1,2,\y,\qb)= }
 \label{gtoy1}\\
&& \left(b_q^{(1)}(q,1,2,\y,\qb) +
b_{\qb}^{(1)}(q,1,2,\y,\qb) + b_3^{(1)}(q,1,2,\y,\qb) \right)
\Big| _{\Nf \to \left(\eL/e_q\right) }  \nn
\eeqn

In order to present the results for the subamplitudes $b^{1\y,(1)}_{12}$
it is convenient to decompose them further.
\beqn
b^{1\y,(1)}_{12}(q,1,2,\y,\qb) &=&
 i \cg  \ \Bc_{12}^{1\y} (q,1,2,\y,\qb)  \label{sadec1} \\
&+& i \cg \Nf \ \Bc_{12}^{1\y, {\rm ext}} (q,1,2,\y,\qb)
+  i \cg \frac{\eL}{e_q} \ \Bc_{12}^{1\y, {\rm int}} (q,1,2,\y,\qb) \nn
\eeqn
In eq. (\ref{sadec1}) we put the loop-coupled diagrams
in $\Bc_{12}^{1\y, {\rm int}}$, whereas the externally-coupled diagrams
are contained in $\Bc_{12}^{1\y, {\rm ext}}$.  Diagrams without a fermion
loop are collected in $\Bc_{12}^{1\y}$. Note that the subamplitude
$b^{1\y,(1)}_{21}$
can be obtained from $b^{1\y,(1)}_{12}$ by simple relabeling
\cite{BDK2q3g,thesis}.

Neglecting for the moment the problems of the coupling of the photon to
quarks of different electric charge, the subamplitude
$b_{12}^{1\y,(1)}(q,1,2,\y,\qb)$ can be constructed  in the same manner as
$b_{\rm tr}^{1 \y,(1)}$. We find
\beqn
\lefteqn{ i \cg  \ \Bc_{12}^{1\y} (q,1,2,\y,\qb)   =} \label{gtoy2} \\
&& \left( b_{123}^{(1)}(q,1,2,\y,\qb) +
b_{132}^{(1)}(q,1,2,\y,\qb) + b_{312}^{(1)}(q,1,2,\y,\qb) +
\Nc b_{3;12}^{(1)}(q,1,2,\y,\qb) \right) \Big| _{\Nf \to  0} \nn
\eeqn

For the remaining needed functions $\Bc_{12}^{1\y,{\rm ext}}$ and
$ \Bc_{12}^{1\y,{\rm int}}$ we will give the explicit results for the different
helicity
configurations, using the notation:
\beq
\Bc (h_1,h_2,h_3) \equiv \Bc (q,+,1,h_1,2,h_2,3,h_3,\qb,-)
\eeq

First of all we remark that it is sufficient to give the subamplitudes for the
helicity configurations\footnote{Since the tree level results vanish for
  the helicity configuration $(+,+,+)$, the corresponding one-loop
  amplitudes are not really needed. We just give them for completeness}
$(-,+,+), (+,-,+), (+,+,-)$ and $(+,+,+)$. The results for all other
helicity configurations can be obtained with the help of discrete
symmetries \cite{BDK2q3g,thesis}.

For the loop-coupled diagrams we find for all helicity configurations
\beq
\Bc_{12}^{1\y,{\rm int}} (h_1,h_2,h_3) =
 - \frac{1}{2} \ \Bc_{\tr}^{1\y, \Nf} (h_1,h_2,h_3) \label{eq:bint}
\eeq
where $\Bc_{\tr}^{1\y, \Nf}$ denotes the part of
$b_{\rm tr}^{1\y,  (1)}$ which is proportional to $ \eL/e_q $.
Furthermore we get
\beq
\Bc_{12}^{1\y,{\rm ext}} (-,+,+) = \Bc_{12}^{1\y,{\rm ext}} (+,-,+) = 0
\eeq
For the other two helicity configurations the externally-coupled diagrams do
not cancel and we obtain
\beq
\Bc_{12}^{1\y,{\rm ext}} (+,+,-) =
  \frac{\B q 1 \B q 2}{3\ \A 1 2 \B \qb 3  \B q 3}
\eeq
and
\beq
\Bc_{12}^{1\y,{\rm ext}} (+,+,+) =
  \frac{\A \qb 1 \A \qb 2 \B 1 2}{3\ \A \qb 3 \A q 3 \A 1 2^2}
  \label{eq:bext}
\eeq
With eq.(\ref{gtoy2}) and eqs.(\ref{eq:bint} -- \ref{eq:bext}) all
ingredients of eq.(\ref{sadec1}) are given.

We turn now to the process $0\to 2q1g 2\y$.
If we start from the process $0\to 2q3g$ and say gluon 2 and gluon 3 become
photons, there is only one color structure left.
\beq
\Ac^{(i)}(q,1,\y_2,\y_3,\qb) =
g \left(\sqrt{2}  e_q\right)^2 \left(\frac{g}{4 \pi}\right)^{2 i} \
(1)_{q \qb} \ b^{2\y,(i)} (q,1,\y_2,\y_3,\qb)
\eeq

The diagrams of this process are again decomposed into
three classes. The first contains diagrams with no quark loop (i.e. we have
to replace $g^2 \to e_q^2$), the second
consists of diagrams where one photon couples to the  quark loop and the
other photon couples to the external quark $(g^2 \to \eL e_q) $,
and finally in the third class
of diagrams both photons couple to the quark loop $ (g^2 \to [e^2]) $. Here
$[e^2]$ denotes $ \sum e_Q^2$. All diagrams of this process
which contain a quark loop and
two photons  coupled to the external quark line, are self-energy insertions on
external lines and vanish therefore in dimensional regularization.

Since the diagrams of the second class cancel each other for all helicity
configurations it is  easy to get the helicity amplitudes.
\beqn
b^{2\y,(1)}(q,1,\y_2,\y_3,\qb) &=&
\bigg[ \sum_{P(1,2,3)} b^{(1)}_{123} (q,1,\y_2,\y_3,\qb) +
\Nc \ b^{(1)}_1 (q,1,\y_2,\y_3,\qb) \label{twophot} \\
&& + \ \ \Nc b^{(1)}_{3;12}(q,1,\y_2,\y_3,\qb)
   + \Nc \ b^{(1)}_{3;21}(q,1,\y_2,\y_3,\qb) \nn \\
&& + \ \Nc \ b^{(1)}_{2;13}(q,1,\y_2,\y_3,\qb)
  + \Nc \ b^{(1)}_{2;31}(q,1,\y_2,\y_3,\qb)
\bigg] \bigg|_{\Nf \to \frac{[e^2]}{e_q^2}} \nn
\eeqn
Obviously, the analytic expression  obtained in this way is rather
massive and further simplification is possible, but we do not
pursue this any longer.

\section{Two external quark pairs}

For processes with external quarks which have different electric charge, the
situation concerning the replacement $g \to e_q$  is more complicated.
Since the gluon couples in the same way to
the different quarks, but the photon does not, it is not possible to get
the amplitudes for the photonic process without additional work.

We denote the charge of $q$
and $Q$ by
$e_q$ and  $e_Q$ respectively. The color decomposition at one loop
$(i = 1)$ and at tree
level $(i = 0)$  reads:
\beqn
\lefteqn{ \Ac^{(i)}(\qb,\Qb;Q,q;\y) = }  \\
&& g^2 \sqrt{2}  \left( \frac{ g}{4 \pi} \right)^{2 i} \bigg[ \
 \del_{q \Qb} \del_{Q \qb} \ e_q u_2^{(i)}(\qb,\Qb;Q,q;\y) +
 \del_{q \Qb} \del_{Q \qb} \ e_Q  d_2^{(i)}(\qb,\Qb;Q,q;\y)
   \nn \\
&&  \qquad  \qquad
- \frac{1}{N} \del_{q \qb} \del_{Q \Qb} \ e_q u_1^{(i)}(\qb,\Qb;Q,q;\y) -
 \frac{1}{N} \del_{q \qb} \del_{Q \Qb} \ e_Q d_1^{(i)}(\qb,\Qb;Q,q;\y)
  \bigg] \nn
\eeqn
Since the two loop-coupled diagrams cancel each other,
the subamplitudes $u^{(1)}$ and $d^{(1)}$ do not
contain terms proportional to $[e]/e_q $.

We write the one-loop subamplitudes  in the following form:
\beqn
 \lefteqn{u_1^{(1)}(\qb,\Qb,Q,q,\y) =} \\
&& i \cg \ u_1^l(\qb,\Qb,Q,q,\y) +
 \frac{i \cg }{\Nc^2} \ u_1^s(\qb,\Qb,Q,q,\y) -
 i \cg  \frac{\Nf}{\Nc} \ u^{\Nf} (\qb,\Qb,Q,q,\y) \nn \\
 \lefteqn{u_2^{(1)}(\qb,\Qb,Q,q,\y) =} \\
&& i \cg  \Nc \ u_2^l(\qb,\Qb,Q,q,\y) +
 \frac{i \cg }{\Nc} \ u_2^s(\qb,\Qb,Q,q,\y)
 + i \cg  \Nf \ u^{\Nf} (\qb,\Qb,Q,q,\y) \nn
 \eeqn
The subamplitudes $d_1^{(1)}$ and $d_2^{(1)}$ are decomposed in the same way.

Since we can use parity to change all helicities, it is sufficient to
consider only helicity configurations with positive photon
helicity. Furthermore, we have the relations
\beq
d_i^{(i)}(\qb,h_{\qb},\Qb,h_{\Qb},Q,h_Q,q,h_q,\y,h_{\y}) =
 u_i^{(i)}(\Qb,h_{\Qb},\qb,h_{\qb},q,h_q,Q,h_Q,\y,h_{\y}) \label{udrel}
\eeq
For the remainder of this section we use the notation
\beq
u^{(i)}(h_q,h_Q,h_\y) \equiv
u^{(i)}(\qb,h_{\qb},\Qb,h_{\Qb},Q,h_Q,q,h_q,\y,h_{\y})
\eeq
We will give the subamplitudes $u^{(i)}(+,+,+)$ and $u^{(i)}(+,-,+)$ below.
The remaining subamplitudes can be obtained either with parity,  eq.
(\ref{udrel})
or
\beqn
u^{(i)}(-,+,+) &=& - u^{(i)}(+,-,+) \big|_{q \lra \qb, Q \lra \Qb} \\
u^{(i)}(-,-,+) &=& - u^{(i)}(+,+,+) \big|_{q \lra \qb, Q \lra \Qb}
\eeqn

First we recall the tree level results,
which can be obtained easily from the tree level results of $0\to 4q1g$
\beqn
 u_1^{(0)}(+,+,+) &=& u_2^{(0)}(+,+,+) =
   - i \ \frac{\A \qb \Qb^2}{\A \Qb Q \A \qb \y \A q \y} \\
 u_1^{(0)}(+,-,+) &=& u_2^{(0)}(+,-,+) =
   i \ \frac{\A \qb Q^2}{\A \Qb Q \A \qb \y \A q \y}
\eeqn

In order to present the results, it is convenient to decompose the
one-loop subamplitudes in the following way:
\beqn
u_1^s(h_q,h_Q,h_\y) &=& u_{A}(h_q,h_Q,h_\y) + u_{B} (h_q,h_Q,h_\y) \\
u_2^s(h_q,h_Q,h_\y) &=& -2\ u_{A}(h_q,h_Q,h_\y) - u_{B}(h_q,h_Q,h_\y) \\
u_2^l(h_q,h_Q,h_\y) &=& u_{A}(h_q,h_Q,h_\y) - u_1^l(h_q,h_Q,h_\y)
\eeqn
So we have to give the results for $u_1^l,u_{A},u_{B}$ and $u^{\Nf}$
for the two helicity configurations $(+,+,+)$ and $(+,-,+)$.
The term proportional to $\Nf$ is the simplest.
\beq
u^{\Nf}(h_q,h_Q,h_\y) = i u_1^{(0)} (h_q,h_Q,h_\y)
 \left( \frac{2}{3} \frac{\P \Qb Q}{\ve} + \frac{10}{9} \right)
\eeq
For the leading terms of $u_1$ we get
\beqn
u_1^l(+,+,+)&=& - i u_1^{(0)}(+,+,+) \left(
 \frac{\P\qb\Qb}{\ve^2} + \frac{\P q Q}{\ve^2} -
 \frac{2}{3} \frac{\P\Qb Q}{\ve} - \frac{29}{18} \right) \\
&+& \frac{\A\qb \y \A q \Qb^2 \B q \y^2}{2 \ \A q \y \A \Qb Q}
              \L2{^{\qb \y}_{\Qb Q}} -
  \frac{\A \qb \Qb \A q \Qb \B q \y}{\A q \y \A \Qb Q} \L1{^{\qb \y}_{\Qb Q}}
   \nn \\
&+& \frac{\A \qb q \A \qb \Qb \B q Q}{2 \ \A \qb \y \A q \y \s \Qb Q } +
 \frac{\A \qb \Qb \A q \Qb \B q \y}{2\ \A q \y \A \Qb Q \s \Qb Q} \nn \\
&-& i u_1^{(0)}(+,+,+) \left( \F (\qb,\Qb,Q) + \F (q,Q,\Qb) +
                            \F (\Qb,\qb,\y) + \F (Q,q,\y) \right) \nn
\eeqn
and
\beqn
u_1^l(+,-,+)&=& - i u_1^{(0)}(+,-,+) \left(
 \frac{\P\qb\Qb}{\ve^2} + \frac{\P q Q}{\ve^2} -
 \frac{2}{3} \frac{\P\Qb Q}{\ve} - \frac{29}{18} \right) \\
&-& \frac{\A \qb \y \A q Q^2 \B q \y^2}{2 \ \ \A q \y \A \Qb Q}
                  \L2{^{\qb \y}_{\Qb Q} } -
\frac{\A \qb q \A \qb \Qb \A q Q \B q \Qb}{\A \qb \y \A q \Qb \A q \y}
                  \L1{^{\qb \y}_{\Qb Q}} -
\frac{\A \qb Q^2}{\A \qb \y \A q \y \A \Qb Q} \L0{^{\qb \y}_{\Qb Q}} \nn \\
&-& \frac{\A \qb \Qb \A \Qb Q \B \Qb \y}{\A q \Qb \A \Qb \y}
     \L1{^{\qb \y}_{q Q}} -
\frac{\A \qb q \A \qb Q}{\A \qb \y \A q \Qb \A q \y} \L0{^{\qb \y}_{q Q}} +
\frac{\A \qb \Qb \A Q \y \B \Qb \y}{\A q \y \A \Qb \y}
      \L1{^{\qb \Qb}_{q Q}} \nn \\
&+& \frac{\A \qb q \A \qb Q \B q \Qb}{2 \ \A \qb \y \A q \y \s \Qb Q} -
\frac{\A \qb Q \A q Q \B q \y}{2 \ \A q \y \A \Qb Q \s \Qb Q} \nn \\
&-& i u_1^{(0)}(+,-,+) \left(  \F (\qb,\Qb,Q) +
 \h \qb q \Qb Q^2 \F (q,Q,\Qb) \right. \nn \\
&& \qquad \qquad \quad
\left.  + \ \h \qb \y \Qb Q^2 \F(\Qb,\qb,\y) + \F (Q,q,\y) \right) \nn
\eeqn
With the help of the relations
\beqn
u_{B} (+,-,+) &=& u_{B} (+,+,+)|_{Q \lra \Qb} \\
u_{A} (+,-,+) &=& - u_{A} (+,+,+)|_{Q \lra \Qb}
\eeqn
and the  functions
\beqn
u_{B}(+,+,+) &=&  i u_1^{(0)}(+,+,+) \left(
 \frac{\P \qb q}{\ve^2} + \frac{\P \Qb Q}{\ve^2} +  \frac{3\ \P \Qb Q}{\ve}
 + \frac{13}{2} \right) \\
&-&  \frac{\A \qb q^2 \A \Qb Q \B q Q^2}{2 \ \A \qb \y \A q \y}
       \L2{^{\Qb Q}_{\qb \y}}
- \frac{\A \qb q \A \Qb \y \B q \y \B Q \y}{\A q \y}
     \L2{^{\qb q}_{\Qb Q}} \nn \\
&+& \frac{\A \qb q \A q \Qb \A \Qb \y \B q \y}{\A q \y^2 \A \Qb Q}
     \left( \L1{^{\Qb Q}_{\qb q}} + \L1{^{\Qb Q}_{\qb \y}} \right) \nn \\
&+& \frac{\A \qb \Qb^2}{2 \ \A \qb \y \A q \y \A \Qb Q}
     \L0{^{\Qb Q}_{\qb \y}} +
\frac{\A \qb q^2 \B q Q^2}{2\ \A \qb \y \A q \y \B \Qb Q \s \qb \y} -
\frac{\A \qb \Qb \B Q \y}{\A q \y \s \Qb Q} \nn \\
&+& i u_1^{(0)}(+,+,+) \left(
 \F (\qb,q,\y) + \h \Qb q \y \qb^2 \F (q,\qb,\y) \right) \nn
\eeqn
and
\beqn
u_{A}(+,+,+) &=& - i u_1^{(0)}(+,+,+) \left(
 \frac{\P \qb \Qb}{\ve^2} + \frac{\P q Q}{\ve^2} -
 \frac{\P \qb Q}{\ve^2} - \frac{\P q \Qb}{\ve^2} \right) \\
&-& \frac{\A \qb q \A q \Qb \B q \y}{\A q Q \A q \y} \L1{^{\qb \y}_{\Qb Q}}
- \frac{\A \qb Q \A \Qb Q \B Q \y}{\A q Q \A Q \y}
           \L1{^{\qb \y}_{q \Qb}} \nn \\
&-& \frac{\A \qb Q \A \Qb \y \B Q \y}{\A q \y \A Q \y} \L1{^{\qb Q}_{q \Qb}}
+ \frac{\A \qb q \A \qb \Qb}{\A \qb \y \A q Q \A q \y} \L0{^{\Qb Q}_{q \Qb}}
\nn \\
&-& i  u_1^{(0)}(+,+,+) \bigg(
 \F(\qb,\Qb,Q) - \h \Qb Q q \qb^2 \F(q,\Qb,Q) + \F(\Qb,\qb,\y)  \nn \\
&& \qquad \qquad \quad  -\ \F(\Qb,q,\y) + \F(\Qb,Q,q) - \F(\Qb,Q,\qb) \nn \\
&& \qquad \qquad  \quad
 + \ \F(Q,q,\y) - \h \Qb Q \y \qb^2 \F (Q,\qb,\y) \bigg) \nn
\eeqn
we obtain all the  needed functions.

\def\np#1#2#3  {{\it Nucl. Phys. }{\bf #1} (19#3) #2}
\def\nc#1#2#3  {{\it Nuovo. Cim. }{\bf #1} (19#3) #2}
\def\pl#1#2#3  {{\it Phys. Lett. }{\bf #1} (19#3) #2}
\def\pr#1#2#3  {{\it Phys. Rev. }{\bf #1} (19#3) #2}
\def\prl#1#2#3  {{\it Phys. Rev. Lett. }{\bf #1} (19#3) #2}
\def\prep#1#2#3{{\it Phys. Rep. }{\bf #1} (19#3) #2}
\def\zp#1#2#3  {Zeit.\ Phys.\ Lett. #1 (19#3) #2}

\end{document}